\documentclass[12pt,a4paper]{article}

\usepackage[left=2cm,right=2cm,top=2.5cm,bottom=2.5cm]{geometry}
\usepackage[utf8]{inputenc}

\usepackage{scalerel,stackengine,amsmath}
\usepackage{amsfonts}
\usepackage{amssymb}
\usepackage{hyphenat}
\usepackage{enumitem}
\usepackage{dsfont}
\usepackage{cite}
\usepackage{hyperref}
\usepackage{graphicx}
\usepackage{caption}
\usepackage{float}
\usepackage{subcaption}
\usepackage{soul}
\usepackage[labelformat=simple]{subcaption}
\usepackage{array}
\usepackage{physics}
\usepackage{float}
\usepackage{mathtools}
\usepackage{rotating}
\usepackage{amssymb}

\newcommand{\citet}{\cite} 
\usepackage{braket}
\usepackage[export]{adjustbox}
\usepackage{mathtools}
\usepackage{comment}
\usepackage{mathrsfs}  

\usepackage{adjustbox}
\usepackage{slashed}
\usepackage{mathrsfs}
\usepackage{lmodern}
\usepackage{bbold}
\usepackage{mmacells}

\numberwithin{equation}{section}
\renewcommand{\thefootnote}{\fnsymbol{footnote}}

\begin{document}

\begin{flushright}
June 2026
\end{flushright}

\begin{center}
{\bf \LARGE{Renormalisation Group Invariants
from\\[3mm] Scaling and Non-overlapping Symmetries} }
\end{center}

\bigskip

\begin{center}
{\large Apostolos Pilaftsis}$\,$\footnote{E-mail address: {\tt
apostolos.pilaftsis@manchester.ac.uk}}\\[3mm] 
{\it Department of Physics
and Astronomy, University of
Manchester,\\ Manchester M13 9PL, United Kingdom}
\end{center}

\bigskip

\centerline{\bf ABSTRACT}
\vspace{2mm}

\noindent
We show how the synergy of scaling and non-overlapping global symmetries can lead to Renormalisation Group Invariants (RGIs) among the parameters of potentials with multiple scalars. The instrumental role of scale-invariant field directions in the identification and construction of RGIs for bilinear field operators to all loops is demonstrated. We present a few illustrative examples to showcase our constructive spurion-field approach, which is applied to simple non-supersymmetric models as well as to scenarios reported recently in the literature that include the two-Higgs doublet model. The problematic issues of RGIs to address the gauge-hierarchy problem beyond supersymmetry are discussed.

\medskip
\noindent
{\small {\sc Keywords:} renormalisation group invariants, scaling symmetries, non-overlapping symmetries,\\ spurion formalism.}

\thispagestyle{empty}

\section{Introduction}\label{sec:intro}

In Quantum Field Theory (QFT), the existence of Renormalisation Group Invariants (RGIs) among the parameters of a theory betray hidden, global or local symmetries in the underlying structure of this theory~\cite{Zimmermann:1984sx}. In this regard, Supersymmetry (SUSY) endows QFTs with powerful\- non-renormalisation theorems that result in RGIs of gauge couplings and all loop-order finiteness with vanishing $\beta$ functions~\cite{Wess:1992,Nilles:1983ge,Haber:1984rc,Martin:1997ns}.  
RGI effective charges~\cite{Brodsky:1982gc,Grunberg:1982fw,Cornwall:1989gv,Binosi:2009qm,Papavassiliou:1997fn,Papavassiliou:1997pb} may also reflect 
physical consequences in  QFTs based on gauge symmetries that are unbroken, such as Quantum\- Electro\-dynamics and Quantum Chromo\-dynamics, or broken via the Higgs mechanism, such as the Standard Model~(SM).

An interesting question that emerges from studying RGIs is whether their accidental presence also signifies the existence of some deeper symmetry within a theoretical model or a given theoretical setting. To put it simply, does an RGI {\em always} have its origin in some known well-defined field-theoretic symmetry? 
In this paper, we show that the answer to this question is not that simple and the origin of an RGI may be the result of a synergy of scaling and non-overlapping global (continuous or discrete) symmetries. In fact, it can happen so that none of these symmetries is an exact symmetry of the theory, but their non-overlapping or partially overlapping action can give rise to an RGI. To address the above question, it is sufficient to consider scalar potentials that contain only two scalar fields. Our analysis is based on the so-called spurion-field formalism and includes the well-studied theoretical framework, known as the Two Higgs Doublet Model~(2HDM)~\cite{Lee:1973iz,Branco:1980sz,Branco:1985aq,Weinberg:1990me,Pilaftsis:1999qt,Darvishi:2023fjh}. Nevertheless, its generalisation to other theoretical frameworks with a significantly extended Higgs sector is straightforward. We also note that our study is motivated by recent articles that address a related topic~\cite{Ferreira:2023dke,Haber:2025cbb,Trautner:2025yxz,deBoer:2025jhc,Trautner:2025prm,Ferreira:2026bex,Grzadkowski:2026gkx}, although our approach and results differ fundamentally from these. As a consequence of our investigations, we find that there are {\em no} new symmetries responsible for generating RGIs in 2HDM, other than those that have already been reported in the literature. 

At this point, it is important to remind ourselves that the 2HDM potential could realise a large number of global symmetries~\cite{Peccei:1977hh,Deshpande:1977rw,Ivanov:2007de,Maniatis:2007vn,Ferreira:2009wh,Battye:2011jj,Pilaftsis:2011ed}, either continuous or discrete. The spontaneous or explicit breaking of these symmetries may result  in  pseudo-Goldstone bosons~\cite{Weinberg:1972fn},    flavour-changing  neutral currents~\cite{Glashow:1976nt,Paschos:1976ay,Pich:2009sp}
and  CP violation~\cite{Branco:1980sz,Branco:1985aq,Botella:1994cs,Darvishi:2023fjh}. In detail, the tree-level 2HDM potential may realise 6 symmetries that preserve U(1)$_Y$~\cite{Ivanov:2007de}, as well as 7 additional symmetries that are custodial and occur when $g' \to 0$~\cite{Battye:2011jj,Pilaftsis:2011ed,BhupalDev:2014bir,Darvishi:2019dbh,Aiko:2020atr}. However, we should caution the reader that some of these global symmetries  require the absence of the Yukawa couplings to allow them to be extended to the complete 2HDM Lagrangian. Finally,
there is one more symmetry, albeit non-compact, if all mass parameters are absent from the 2HDM potential~\cite{Lee:2012jn}. 
In this case,  the theory becomes Scale Invariant~(SI), and going beyond the tree level is essential to obtain a phenomenologically viable and testable model~\cite{Lee:2012jn,Eichten:2022vys,Slavich:2026zmi}.

Since our analysis relies on the spurion-field formalism by exploiting global symmetries of the 2HDM, we do not provide explicit expressions for all relevant $\beta$-functions. Nevertheless, the validity of our results has been verified against existing results in the literature, including~\cite{Carena:2015uoe,Bednyakov:2018cmx,Ferreira:2023dke}.
Furthermore, we expect that other related methods which make use of the so-called {\em reduction equations} of the 2HDM~\cite{Pech:2023bjm} will be compatible with our findings. 

After this introductory section, the paper exhibits the following organisation.  Section~\ref{sec:Sym} introduces the spurion formalism, as well as defines the notion of scaling and non-overlapping symmetries. In the same section, we also give simple examples to elucidate their application in minimal theoretical settings. In Section~\ref{sec:RGIs}, we consider a few illustrative scenarios of a U(1) model with two complex scalar fields, where the synergy of scaling and non-overlapping symmetries in obtaining RGIs is demonstrated by employing the spurion formalism. Based on these findings, we turn our attention to 2HDM in Section~\ref{sec:2HDM}. In particular, we show how the identification of an SI field direction enables us to construct RGIs for all dimensionful mass parameters of the 2HDM potential. In fact, we find a separation of all its mass scales under RG running in an approximate CP2-symmetric 2HDM and discuss its relevance to address the gauge hierarchy problem beyond SUSY. 
Finally, Section~\ref{sec:concl} gives a succinct summary of the present study and presents future research directions for further exploration.

\renewcommand{\thefootnote}{\arabic{footnote}}
\section{Scaling and Non-overlapping Symmetries}\label{sec:Sym}

In our considerations, we assume that the classical action $S$ describing the theory has a generic form
\begin{equation}
    \label{eq:Saction}
S[\phi ; {\rm x} ]\: \equiv\: \int\!\text{d}^4 x\, {\cal L}[\phi; {\rm x}]\: =\: \int\!\text{d}^4 x\, \Big(\partial_\mu\phi^*\, \partial^\mu\phi\: +\: 
V(\phi;{\rm x} )\: +\: \dots\Big)\,.     
\end{equation}
In the above, ${\cal L}[\phi;{\rm x}]$ is the Lagrangian, $\phi$ represents a generic field or a collection of fields, e.g.~$\phi = \{\phi_i\}$, that transform under a global group $G$, the variable ${\rm x}$ runs over all kinematic parameters, such as masses and couplings, $V(\phi;{\rm x})$ is the scalar potential invariant under the action of $G$, and the ellipses indicate the possible existence of fermions and gauge bosons in the theory, should $G$ be gauged. In addition, the Lagrangian ${\cal L}[\phi;{\rm x}]$ is assumed to be renormalisable containing operators up to energy dimension 4. 

A theory is said to be invariant under symmetry transformations described by the group~$G$, i.e.~under 
$\phi \to \phi' = G\!\cdot\!\phi$, if its classical action transforms as
\begin{equation}
    \label{eq:SGinvariant}
S[\phi ; {\rm x} ]\ \to\ S[\phi' ; {\rm x} ]\: =\: S[G\!\cdot\!\phi ; {\rm x} ]\: =\: S[\phi ; {\rm x} ]\,.   
\end{equation}
If the theory is {\em not} invariant under $G$, there is often the possibility to make it invariant by treating the kinematic parameters ${\rm x}$ as external spurious fields, also known as spurions, possessing compensating charges. In this case, one may formally expect that
\begin{equation}
    \label{eq:SGspurions}
S[\phi ; {\rm x} ]\ \to\ S[\phi' ; {\rm x}' ]\: =\: S[G\!\cdot\!\phi ; G\!\cdot\!{\rm x} ]\: =\: S[\phi ; {\rm x} ]\,.   
\end{equation}
This is the essence of spurion techniques used in the literature~\cite{Ellis:1978hq,Craig:2021ksw,deLima:2024hnk,deLima:2024lfc} to determine the para\-metric structure of UV divergences, as well as the expected form of RG equations. To further facilitate the construction or identification of RGIs, we must introduce two mathematical symmetries or concepts: (i)~non-overlapping symmetries and (ii)~scaling or dilatonic symmetries.

\subsection{Non-overalapping Symmetries}

The first concept refers to {\em non-overlapping symmetries}.   Non-overlapping symmetries are pos\-sible symmetries of the theory that are described by compact groups which may be discrete or continuous and do not share the same generators or have no generators in common. For example, consider~$\mathbb{Z}_6 =\{\pm 1\,,\pm \omega\,, \pm \omega^2\}$ with $\omega = e^{\pi i/3}$ and $\mathbb{Z}_4 = \{\pm 1\,, \pm i\} $.
Then, the intersection of the two groups is 
\begin{equation}
   \label{eq:Z4capZ6}
\mathbb{Z}_6 \cap \mathbb{Z}_4\: =\: \mathbb{Z}_2 = \{ \pm 1\}\,,  
\end{equation}
which is a much smaller group, while $\mathbb{Z}_4 \not\subset \mathbb{Z}_6$. A non-overlapping group is sometimes referred to as a partially overlapping group if there is a partial non-trivial overlap between the two groups, as happens in the example above, since $\mathbb{Z}_2 \ne \mathbb{I} =\{ 1\}$. 

To give a concrete field-theory example based on the groups appearing in~\eqref{eq:Z4capZ6}, we  consider a model with two complex scalar fields $\phi_i$ (with $i=1,2$), whose potential is holomorphic and reads
\begin{equation}
    \label{eq:VZ6-Z4}
V(\phi_i)\: =\: V_{\mathbb{Z}_6}(\phi_i)\: +\: V_{\mathbb{Z}_4}(\phi_i)\,, 
\end{equation}
with~\footnote{Note that only terms with even powers of fields contribute to the potential $V(\phi_i )$, and so $V(\phi_i )$ has an extra $\mathbb{Z}_2$ discrete symmetry which we ignore in this discussion.}$^{,}$\footnote{The imposition of $\mathbb{Z}_4$ on the Lagrangian does not forbid the holomorphic kinetic mixing term $\zeta_{12}\,(\partial_\mu \phi_1)( \partial^\mu \phi_2)$, in addition to the non-holomorphic kinetic terms $(\partial_\mu \phi^*_i)( \partial^\mu \phi_i)$, with $i=1,2$. However, if one sets $\zeta_{12} =0$ in the Born approximation, no non-zero $\zeta_{12}$ will be generated perturbatively at higher orders through holomorphic scalar potential interactions only.}
\begin{eqnarray}
   \label{eq:VZ6}
V_{\mathbb{Z}_6}(\phi_i) \!&=&\!  \lambda_{12}\,\phi^2_1\phi^2_2\ +\  {\rm H.c.},\\[2mm]
   \label{eq:VZ4}
V_{\mathbb{Z}_4}(\phi_i) \!&=&\! m^2_{12}\,\phi_1\phi_2\: +\: \lambda_1\, \phi_1^4 \: +\: \lambda_2\, \phi_2^4\: +\:  \lambda'_{12}\,\phi^2_1\phi^2_2\ +\ {\rm H.c.}\qquad
\end{eqnarray}
In writing the above potentials, the following discrete charge assignments were made for the fields under the discrete group $\mathbb{Z}_6$ ($\mathbb{Z}_4$): 
\begin{equation}
  \label{eq:Z6Z4charges}
\phi_1:\ \omega\: (+i)\,,\qquad \phi_2:\ \omega^2\: (-i)\,.
\end{equation}
Notice that the union $\mathbb{Z}_6\cup \mathbb{Z}_4$ is not a group under the law of multiplication, since the elements $\pm i\omega,\, \pm i\omega^2$ are not in the union. 

By switching on and off different kinematic parameters in the potential~\eqref{eq:VZ4}, two very distinct non-overlapping symmetries can be realised in the theory. The individual terms of $V_{\mathbb{Z}_4}$ break $\mathbb{Z}_6$, which carry the following 
compensating charges:
\begin{equation}
    \label{eq:Z6Z4spurion}
m^2_{12}:\ \omega^3 = -1\,,\quad \lambda_1:\ \omega^2\,,\quad \lambda_2:\ \omega^4 = -\omega\,,\quad \lambda'_{12}:\ \omega^0 =+1\, .      
\end{equation}
Observe that $\lambda'_{12}$ does not carry compensating charge and therefore is neutral under $\mathbb{Z}_6$, that is, $\lambda'_{12}$ is a $\mathbb{Z}_6$ invariant. As we will elucidate in the next section while discussing concrete models, such kinematic parameters will  play the role of spurions, which help to control the UV properties of the theory. 

Other examples relevant to 2HDMs are the global groups  $\text{CP2}$ and $\text{U}(1)_{\rm PQ}$, which have a similar partial overlap, i.e.~$\text{CP2} \cap \text{U}(1)_{\rm PQ} = \mathbb{Z}_2$, since $(\text{CP2})^2 = -\mathbb{1}_2$, where $\mathbb{1}_2$ is the 2D identity matrix. We will discuss such scenarios in Sections~\ref{sec:RGIs} and~\ref{sec:2HDM}.

\subsection{Scaling Symmetries}

The second class of symmetries within a theory refers to the {\em classical scaling symmetries}. If we set all mass terms to zero in the bare Lagrangian, including those present in the scalar potential, the classical action $S$ in~\eqref{eq:Saction} becomes SI. This means that under a scaling or dilatonic change of the spacetime coordinates, $x' = e^\epsilon x$ (with $\epsilon \in \mathbb{R}$), a generic field $\phi(x)$ transforms as
\begin{equation}
    \label{eq:SItransf}
\phi (x)\ \to\  \phi'(x')\: =\: e^{\epsilon a}\, \phi(e^\epsilon x)\,,    
\end{equation}
leading to $S'[\phi'] \equiv S[\phi'] = S[\phi]$, where $a$ is the scaling dimension of the field~$\phi(x)$~\cite{Gildener:1976ih,Alexander-Nunneley:2010tyr}. At tree~level, the field~$\phi(x)$ takes the value $a=1\: (3/2)$ for a boson (fermion). 

This second class of symmetries, although not protected at the quantum level, plays an important role in efficiently identifying RGIs in the Dimensional Regularisation (DR) scheme~\cite{tHooft:1973mfk}. It~enforces the absence of a combination of dimensionful parameters along a specified SI field  direction, which can be valid to all loop orders if the scaling symmetry of this field direction is not violated by other interactions in the theory. Consequently, a sufficient condition for constructing dimensionful RGI bilinears is the existence of SI field directions for a given theory of interest. This is what we intend to exemplify in the next section by analyzing some illustrative models.

\section{Renormalisation Group Invariants}\label{sec:RGIs}

As mentioned in the previous section, classical scaling symmetries are not protected at the quantum level, but violated by global Weyl anomalies. As happens with chiral global anomalies, the emergent scaling-violating terms that arise from quantum loops do not need renormalization. This in turn implies that these scaling-violating terms do not have overall UV divergences, but only at the subgraph level which were removed by the remaining counter-terms (CTs) of the theory, like scalar quartic, gauge and Yukawa couplings. 

An archetypal example of an SI model is the SM, {\em without} the bilinear mass term, $-\mu^2 \Phi^+\Phi$, in the scalar potential~\cite{Coleman:1973jx}:
\begin{equation}
    \label{eq:VSM}
 V(\Phi )\; =\; \frac{\lambda}{2}\,(\Phi^+\Phi)^2\,, 
\end{equation}
where $\Phi$ is the Higgs doublet and $\lambda$ is a non-negative scalar quartic coupling. As a consequence of the classical scaling symmetry, the following {\em master} Ward Identity (WI) can be derived from the classical action~\cite{Gildener:1976ih,Alexander-Nunneley:2010tyr}:
\begin{equation}
    \label{eq:WIscale}
 \frac{\partial V(\Phi )}{\partial \Phi}\,\Phi\: +\: \Phi^\dagger\, \frac{\partial V(\Phi )}{\partial \Phi^\dagger} \; =\; 4\,V(\Phi)\, .
\end{equation}
Beyond the tree level, the above WI is broken in two ways: (i)~by global Weyl anomalies which are UV-finite, non-renormalizable logarithmic terms of the form $(\Phi^+\Phi)^2\ln (\Phi^\dagger \Phi/v^2_\Phi)$, where ${v^2_\Phi = \langle \Phi^\dagger \Phi \rangle}$ is the squared SM vacuum expectation value (VEV), and (ii)~explicitly by employing dimensionful regulators to isolate the UV divergences, e.g.~as done within the framework of a $\Lambda$-cutoff regularisation 
scheme~\cite{Branchina:2022jqc}. 

If a $\Lambda$-dependent regularisation scheme is adopted, one then has to restore the classical scaling symmetry of the {\em local classical action} at the
quantum level. This means that all possible evanescent scale-violating operators, such as $c_\Lambda\,\Lambda^4$ and $c_\Phi\,\Lambda^2\, \Phi^\dagger \Phi$, should be added to $V(\Phi )$ in~\eqref{eq:VSM}, consistent with SM gauge invariance. The coefficients   $c_{\Lambda,\Phi}$ are then determined from the renormalisation conditions: 
\begin{equation}
  \label{eq:WIren}
\Big(V(\Phi) \, +\, V^{(1),\text{UV}}(\Phi ) \Big)\Big|_{\Phi = 0} =\: 0\,, \qquad
\frac{\partial^2 \big(V(\Phi) \, +\, V^{(1),\text{UV}}(\Phi)\big)}{\partial\Phi^\dagger\, \partial\Phi}\bigg|_{\Phi = 0} = 0\,,    
\end{equation}
where~$V^{(1),\text{UV}}(\Phi )$ stands for the UV-divergent part of the unrenormalised one-loop effective potential\-~$V^{(1)}(\Phi )$. It is important to note here that the two conditions in~\eqref{eq:WIren} are a direct consequence of the master WI~\eqref{eq:WIscale}, by requiring that the local UV-infinite part of the theory preserves the classical scale  invariance. In fact, the second condition in~\eqref{eq:WIren} ensures the absence of quadratic divergences proportional to $\Lambda^2$, and as such, classical scaling symmetry may be regarded as a valid alternative of renormalisable theories, along with supersymmetry, to technically solve the infamous gauge hierarchy problem~\cite{Gildener:1976ai}.  

In the DR scheme, the two conditions~\eqref{eq:WIren} are automatically fulfilled for   
$V^{(1),\text{UV}}(\Phi )$, implying the vanishing of the symmetry-restoring CTs, i.e.~$c_{\Lambda,\Phi}=0$ to all orders. For this reason, we adopt the DR scheme in this work. Evidently, scaling symmetries are essential to determine the RG flow of the theoretical parameters. For the case of the SM, we obtain an obvious RG fixed point as $\mu^2 \to 0$, i.e.
\begin{equation}
   \label{eq:RGSMmu2}
\beta_{\mu^2}\big|_{\mu^2\to 0}\equiv \frac{{\rm d}\mu^2}{{\rm d}t}\bigg|_{\mu^2\to 0}\ =\ 0\,,    
\end{equation}
with $t \equiv \ln Q^2/Q^2_0$, where $Q\, (Q_0)$ is an RG (reference) scale. As a consequence of~\eqref{eq:RGSMmu2}, the bilinear mass parameter $\mu^2$ is multiplicative renormalisable within the SM, which implies $d\mu^2/dt \propto \mu^2$. We will demonstrate below how scaling symmetries facilitate our quest to efficiently find RGI fixed points in a given theory.

\subsection{Global U(1) Model}\label{subsec:SimpleModel}

Let us discuss a more instructive model that features two complex scalar fields, $\phi_1$ and~$\phi_2$. 
The model posesses a global U(1) symmetry, where both $\phi_1$ and $\phi_2$ carry the same global U(1) charge, e.g.~$+1$. Thus, its scalar potential reads~\footnote{Our notation of parameters followed here is motivated by that of the 2HDM to be discussed in Section~\ref{sec:2HDM}.}
\begin{eqnarray}
   \label{eq:VU1}
V(\phi_1\,,\phi_2) \!& =&\!   m^2_{11}\, |\phi_1|^2\: +\: m^2_{22}\, |\phi_2 |^2\: +\:
\big( m^2_{12}\, \phi^*_1 \phi_2 + {\rm c.c.}\big)\: +\: 
\lambda_1 |\phi_1|^4\: +\: 
\lambda_2 |\phi_2|^4\: +\: \lambda_3 |\phi_1 |^2|\phi_2|^2\nonumber\\
\!&&\! +\, \big[ \lambda_5\,(\phi^*_1 \phi_2)^2 + {\rm c.c.} \big]\: +\, \big(\lambda_6\,|\phi_1|^2 \phi^*_1 \phi_2 + {\rm c.c.} \big)\: +\: \big( \lambda_7\, |\phi_2|^2 \phi^*_1 \phi_2 + {\rm c.c.} \big)\: .
\end{eqnarray}
The potential of the global U(1) model has 4 real mass parameters, $m^2_{11}, m^2_{22},\, \text{Re}\,m^2_{12}$ and~$\text{Im}\,m^2_{12}$, and 9 real quartic couplings, $\lambda_{1,2,3},\, \text{Re}(\lambda_{5,6,7})$ and $\text{Im}(\lambda_{5,6,7})$. It should be noted that a 2D rotation $\mathbb{O}(2)$ on the field basis $(\phi_1,\phi_2)^{\sf T}$ can be performed to go to another basis in which $\lambda_6 = \lambda_7$. Moreover, $\text{Im}(\lambda_5)$ can be removed by rephasing $\phi_2$, without spoiling the constraint $\lambda_6 = \lambda_7$. In the end, of 13 parameters, only 10 of them will be physical if no other interactions are present in the theory.

\subsubsection{The Peccei--Quinn symmetric limit}

In the absence of the parameters
$m^2_{12}$ and $\lambda_{6,7}$, the scalar potential in~\eqref{eq:VU1}, denoted here as~$V_{\rm PQ}$, has an extra Peccei--Quinn (PQ) symmetry~\cite{Peccei:1977hh} U(1)$_{\rm PQ}$, in addition to global~U(1), where the field $\phi_1$ carries the PQ charge: $+1$ and the field $\phi_2$: $-1$. As a consequence, the PQ-symmetric scalar potential obeys the relation: 
\begin{equation}
    \label{eq:VPQid}
 V_{\rm PQ}(\phi_1\,,\phi_2)\: =\: V_{\rm PQ}(e^{i\theta}\phi_1\,, e^{-i\theta}\phi_2)\,,   
\end{equation}
beyond the 
relation: $V_{\rm PQ}(\phi_1\,,\phi_2)\, =\, V_{\rm PQ}(e^{i\chi}\phi_1\,, e^{i\chi}\phi_2)$, due to the original global U(1) symmetry${}$, where~$\theta$ and $\chi$ are arbitrary real constants. It therefore proves more instructive to discuss the PQ symmetric limit first before embarking on the most general case. Due to the PQ symmetry, the~parameters
$m^2_{12}$ and $\lambda_{5,6,7}$ will remain absent to all orders in perturbation theory, i.e.~$m^2_{12}=0$ and $\lambda_{5,6,7}=0$. 

One may now wonder whether such radiatively stable and RGI fixed points could be possible for a combination of other kinematic parameters of the scalar potential $V_{\rm PQ}$ in other symmetric limits of the present theory besides the PQ limit. Obviously, the model exhibits a classical scaling symmetry
in the vanishing limit of the dimensionful parameters $m^2_{11,22}$, and as such, ${\rm d}m^2_{11,22}/{\rm d}t \to 0$, as $m^2_{11,22}\to 0$, according to our discussion in Section~\ref{sec:Sym} [also cf.~\eqref{eq:RGSMmu2}]. 

To find another realisation of a possible classical scaling symmetry, one has to look for~SI field directions in the scalar potential. It is then not difficult to see that 
the potential,
\begin{equation}
    \label{eq:VSIPQ}
V_{\rm SIPQ}(\phi_1\,,\phi_2)\: =\:    m^2_{11}\, \big(|\phi_1|^2\: -\: |\phi_2 |^2\big)\: +\: 
\lambda_1 |\phi_1|^4\: +\: 
\lambda_2 |\phi_2|^4\: +\: \lambda_3 |\phi_1 |^2|\phi_2|^2\,,  
\end{equation}
is SI in the field direction: 
\begin{equation}
   \label{eq:sigma12}
|\phi_1|\: =\: |\phi_2|\: \ne\: 0\;,   
\end{equation}
iff one sets 
\begin{equation}
   \label{eq:m2-m1}
m^2_{22}\: =\: -\,m^2_{11}\;\qquad \text{or}\qquad m^2_{11}\,+\, m^2_{22}\: =\: 0   
\end{equation}
in the PQ-symmetric potential~$V_{\rm PQ}$. To promote this symmetry to all
orders, it is therefore crucial that the squared mass restriction~\eqref{eq:m2-m1} is
reinforced by a symmetry other than U(1)$_{\rm PQ}$. Specifically, we need 
a {\em non-overlapping} symmetry, such that, under its action, 
the expression $|\phi_1| -|\phi_2|$ is odd, i.e.
\begin{equation}
    \label{eq:sigma12odd}
    |\phi_1| - |\phi_2|\ \to\ -\,\big(|\phi_1| - |\phi_2|\big)\,.
\end{equation}
Luckily, as already mentioned in Section~\ref{sec:Sym}, such a {\em non-overlapping} symmetry exists, which is the analog of CP2 in this model. Its action on the
fields $\phi_{1,2}$ is given by
\begin{equation}
    \label{eq:CP2U1}
\text{CP2:}\quad    \phi_1\ \to\ \phi'_1\, =\, \phi^*_2\,,\qquad \phi_2\ \to\ \phi'_2\, = -\phi^*_1\,.
\end{equation}
Upon imposing CP2 on $V_{\rm SIPQ}$, one finds the additional constraint on the quartic couplings:
\begin{equation}
  \label{eq:lambda1=2}
  \lambda_1\: =\: \lambda_2\, ,
\end{equation}
which in turn implies a potential restricted by the form,
\begin{equation}
    \label{eq:VSIPQCP2}
\widetilde{V}_{\rm SIPQ}(\phi_1\,,\phi_2)\: =\:    m^2_{11}\, \big(|\phi_1|^2\: -\: |\phi_2 |^2\big)\: +\: 
\lambda_1 \big(|\phi_1|^4\: +\: |\phi_2|^4\big)\: +\: 
\lambda_3\, |\phi_1 |^2|\phi_2|^2\,.     
\end{equation}
Observe that, in addition to SI and PQ invariance, $\widetilde{V}_{\rm SIPQ}$ is exactly CP2 invariant along the field-space ray: $|\phi_1| = |\phi_2|$. Likewise, the kinetic terms for the complex fields $\phi_{1,2}$ are also invariant under the action of the SI, PQ and CP2 groups. Hence, the anomalous dimensions of the fields~$\phi_{1,2}$, $\gamma_{\phi_{1,2}}$, remain equal to all orders:
\begin{equation}
   \label{eq:gsigma1=2}
 \gamma_{\phi_1}\ =\ \gamma_{\phi_2}\; .   
\end{equation}
If fermions are added to the theory and their SI Yukawa interactions are PQ and CP2 invariant, then the property in~\eqref{eq:gsigma1=2} will continue to hold. 

If SI were exactly preserved beyond the Born approximation along the field ray: ${|\phi_1| = |\phi_2|}$, then no operator linear to $r^0\equiv |\phi_1|^2 +|\phi_2|^2$ would be generated to all loop orders. Given that SI is anomalously broken by UV-finite effects of ${\cal O}[(r^0)^3]$ and higher, the above conclusion is still true, which means that no UV-infinite or UV-finite term $\propto r^0$ will appear at the quantum level in the DR scheme.
Consequently, we find that the combination $m^2_{11} + m^2_{22}$ is a RGI quantity:
\begin{equation}
    \label{eq:RGIm12}
\beta_{m^2_{11} +m^2_{22}}\Big|_{m^2_{22}\to -m^2_{11}}\ =\ 0\,,
\end{equation}
within the PQ-invariant model with the CP2 constraint $\lambda_1 = \lambda_2$, which was applied only to the SI part of the theory. As a byproduct of this analysis, we also have: $\beta_{\lambda_1} = \beta_{\lambda_2}$, or $\beta_{\lambda_1-\lambda_2} = 0$. 

The above discussion might give the impression that there may exist some unconventional (unphysical) symmetry that governs the entire theory for which odd powers of the operator $r^0\equiv |\phi_1|^2 +|\phi_2|^2 > 0$ are disallowed, e.g.~by taking $r^0\to -r^0$~\cite{Ferreira:2023dke}. However, the vanishing of an operator $\propto r^0$ in the scalar potential does not exclude the possibility of generating operators of dimension-6 and higher, such as UV-finite terms ${\propto (r^0)^3/m^2_{11}\,, (r^0)^5/(m^2_{11})^3, \dots}$, when computing the one-loop effective potential in a field bilinear covariant manner~\cite{Pilaftsis:2024uub}. Similar observations can be made for the 2HDM which will be discussed in the next section.

\subsubsection{Beyond the PQ symmetry}

Let us consider the general scalar potential $V$ of the U(1) model given in~\eqref{eq:VU1}, with the aim of identifying RGI quantities beyond those found due to the PQ symmetry. To this end, we proceed as in the previous subsection and require that the  potential $V$ be SI along the complex field directions~\footnote{Note that under the CP2 transformations~\eqref{eq:CP2U1}, we have: $\phi_{1,2+} \to \phi_{1,2-}$, with $\phi_{2+} = -\phi_{2-}$ and $\phi_{1+} = \phi_{1-}$.}
\begin{equation}
   \label{eq:sigmadir} 
\phi_{2\pm}\: =\: \pm\, i\, e^{-i\theta_3}\,\phi_{1\pm}\,,
\end{equation}
with $\theta_3 = {\rm arg}(m^2_{12})$. Explicitly, the bilinear part of the scalar potential~\eqref{eq:VU1} takes the form of
\begin{equation}
    \label{eq:VU1m2}
V_{m^2}(\phi_1\,,\phi_2)\ =\ m^2_{11}\, \big(|\phi_1|^2\, -\, |\phi_2 |^2\big)\: +\: |m^2_{12}|\,
\big( e^{i\theta_3} \phi^*_1 \phi_2 + e^{-i\theta_3} \phi^*_2\phi_1\big)\, ,   
\end{equation}
which implies
\begin{equation}
   \label{eq:VU1constr}
m^2_{11} + m^2_{22}\: =\: 0\quad \mbox{and}\quad  m^2_{12}\: \neq\: 0\,. 
\end{equation}
In other words, along the field directions~\eqref{eq:sigmadir},  the parameter $m^2_{12} \in \mathbb{C}$ remains unconstrained by scale invariance.

\begin{table}[t!]
    \centering
    \begin{tabular}{|l||c c c c c c c c c|}
        \hline
Symmetry & ~$m^2_{11}$~ &  ~$m^2_{22}$~ & ~$m^2_{12}$~ & ~$\lambda_1$~ & ~$\lambda_2$~ & ~$\lambda_3$~ & ~$\lambda_5$~ & ~$\lambda_6$~ & ~$\lambda_7$~ \\[0.1cm]
        \hline\hline
        U(1)$_{\rm PQ}$ & $-$ & $-$ & 0 & $-$ & $-$ & $-$ & 0 & 0 & 0 \\
        \hline
        CP2 & $-$ & $m^2_{11}$ & 0 & $-$ & $\lambda_1$ & $-$ & $-$ & $-$  & $-\lambda_6$\\
        \hline
    \end{tabular}
\caption{\em U(1)$_{\rm PQ}$ and CP2 symmetry relations of kinematic parameters in the global U(1) model with two complex scalar fields.}
    \label{tab:U1Param}   
\end{table}

To protect the field direction $|\phi_1|=|\phi_2|$ from UV-divergent radiative corrections that could be orthogonal to it and spoil this equality [cf.~\eqref{eq:sigma12}], we impose the CP2 symmetry~\eqref{eq:CP2U1} on the SI dimension-4 part of the scalar potential~\eqref{eq:VU1}. As shown in Table~\ref{tab:U1Param}, 
the following constraints are obtained:
\begin{equation}
    \label{eq:lambda1267}
\lambda_1\: =\: \lambda_2\,, \qquad \lambda_6\: =\: -\,\lambda_7\,,\qquad \lambda_5\: \ne\: 0\,.
\end{equation}
Since the scalar potential term $V_{m^2}$ vanishes identically along the two complex field rays~\eqref{eq:sigmadir}, the scalar potential is exactly CP2 invariant in this direction. In the field basis in which $\lambda_6 = \lambda_7$~\cite{Davidson:2005cw,Maniatis:2011qu}, the second constraint in~\eqref{eq:lambda1267} would imply $\lambda_6 = \lambda_7 = 0$.

\begin{table}[t!]
    \centering
    \begin{tabular}{|c||c c  |}
        \hline
Fields & PQ charge & CP2 parity \\[1mm]
        \hline\hline
$\phi_1$ & $+1$ &   \\
        \hline
$\phi_2$ & $-1$ &   \\
\hline
$|\phi_1| \pm |\phi_2|$ & $0$ & $\pm 1$   \\
\hline
$\phi^*_1\phi_2\: (\phi^*_2\phi_1)$ & $-2\: (+2)$ & $-1$\\
        \hline\hline
Spurions & PQ charge & CP2 parity \\[1mm]
\hline\hline
$m^2_{11} \pm m^2_{22}$ & $0$ & $\pm 1$   \\
\hline
$m^2_{12}\, (m^{2\,*}_{12})$ & $+2\: (-2)$ & $-1$   \\
\hline
$\lambda_1\pm \lambda_2$ & $0$ & $\pm 1$   \\
\hline
$\lambda_3$ & $0$ & $+1$   \\
\hline
$\lambda_5\, (\lambda^*_5 )$ & $+4\: (-4)$ & $+1$   \\
\hline
$\lambda_6\pm \lambda_7$ & $+2$ & $\mp 1$   \\
\hline
$\lambda^*_6\pm \lambda^*_7$ & $-2$ & $\mp 1$   \\
\hline
    \end{tabular}
 \caption{\em PQ and CP2 charges of fields or field bilinears, and the associated spurion charges of parameter expressions in the U(1) model with two complex scalars. }
    \label{tab:U1spurion}
\end{table}

In the field basis $\lambda_6 = \lambda_7 = 0$, one might expect a mixing in the renormalisation and in their respective RGEs among the parameters ${m^2_{11}+m^2_{22}}$, ${m^2_{11}-m^2_{22}}$ and $m^2_{12}$. Under PQ transformations, ${m^2_{11}\pm m^2_{22}}$ remain invariant, i.e.~they are U(1)$_{\rm PQ}$ singlet spurions, without PQ charge, or its spurion PQ charge will be zero. Instead, $m^2_{12}$ carries a spurion PQ charge $+2$. The quartic couplings $\lambda_{1,2,3}$ do not carry a spurion PQ charge, but $\lambda_5$ has a spurion PQ charge~$+4$. Table~\ref{tab:U1spurion} gives all the (spurion) PQ charges of the fields (parameters), including CP2 spurion charges, in a general field basis. 

The above considerations imply that the beta-function $\beta_{m^2_{11}+m^2_{22}}$ has no spurion PQ charge, while $\beta_{m^2_{12}}$ has a spurion PQ charge: $+2$. Evidently, no terms linear in $m^2_{12}$ can appear in $\beta_{m^2_{11}+m^2_{22}}$, while terms $\propto |m^2_{12}|^2$ and higher powers, with zero charges, are UV finite and do not contribute to the RG equations. Hence, for the calculation of $\beta_{m^2_{11}+m^2_{22}}$, we get the same result, as if we were going to put $m^2_{12}$ to zero in the field basis $\lambda_6 = \lambda_7 = 0$. Consequently, we obtain the same results as in the PQ symmetric limit of the theory, but without restrictions on $m^2_{12}$ and $\lambda_5$.

Now, one may be concerned that, with the potential addition of fermions, the particular field basis $\lambda_6 = \lambda_7 = 0$ may obscure the parameter relations in the Yukawa sector. This could be an issue if the Yukawa interactions obey an exact CP2 symmetry in a more general field basis with $\lambda_6 = \lambda_7 \ne 0$. In this case, we have to proceed differently. Under CP2, $m^2_{11}+m^2_{22}$ is even, whereas $m^2_{11}-m^2_{22}$ and $m^2_{12}$ are odd if treated as spurions. Moreover, the quartics: $\lambda_1 +(-) \lambda_2$, $\lambda_{3,5}$, and $\lambda_6 -(+)\lambda_7$, are even~(odd) [cf.~Table~\ref{tab:U1spurion}]. On naive dimensional grounds, the following expansion for the beta-function $\beta_{m^2_{11}+m^2_{22}}$ may be written:
\begin{equation}
    \label{eq:betaU1}
\beta_{m^2_{11}+m^2_{22}}\: =\: A\, (m^2_{11} + m^2_{22})\, +\, B\, (m^2_{11} - m^2_{22})\, +\, C\, m^2_{12}\, +\, C^* \, m^{2\,*}_{12}\; .
\end{equation}
To all orders in perturbation theory, the coefficients $A,B$ and $C$ can only be polynomials of dimensionless couplings, such as quartic, Yukawa and gauge couplings. Restricting ourselves to quartics only~\footnote{The addition of fermions to this theory is more involved and will be discussed within the context of the 2HDM in Section~\ref{sec:2HDM}. In the same section (see Subsection~\ref{subsec:SUSY}), we give more details about the derivation of~\eqref{eq:betaU1}.}, we see that $A$ has a spurion CP2 charge of~$+1$, whereas $B$, $C$ and $C^*$ must have
a corresponding charge of~$-1$. Consequently, the coupling polynomials $B$, $C$ and $C^*$ must be proportional to $\lambda_1 - \lambda_2$ or $\lambda_6 +\lambda_7$ to produce an expression for $\beta_{m^2_{11}+m^2_{22}}$ with even spurion CP2 charge, i.e.~CP2-parity $+1$. However, under the assumed restrictions~\eqref{eq:lambda1267}, the coefficients~$B$, $C$ and $C^*$ will vanish. Hence, we find $\beta_{m^2_{11}+m^2_{22}} = A\, (m^2_{11} + m^2_{22})$, which vanishes when the constraint~\eqref{eq:VU1constr} is implemented. 

Finally, we should comment that the global U(1) symmetry of the present model could have been gauged without altering the above conclusion that $m^2_{11}+m^2_{22}$ is an RGI quantity with  $\beta_{m^2_{11}+m^2_{22}} = 0$  when ${m^2_{22} = -m^2_{11}}$, but for arbitrary complex values of $m^2_{12}$, $\lambda_5$ and $\lambda_6 = -\lambda_7$. The discussion in this subsection has highlighted the fact that conventional non-overlapping symmetries exist, such as scale invariance, CP2 and PQ symmetries, which can explain the emergence of RGI expressions in certain symmetric limits of a theory that include SI field directions. In Section~\ref{sec:2HDM}, we will employ the same strategy to unravel the presence of similar RGIs in the 2HDM and comment on the difference of our approach to supersymmetric models.

\subsubsection{Other Scale-Invariant field directions}

Given that scale invariance plays an instrumental role in identifying RGIs, we may investigate whether other SI field directions exist in the global U(1) model that are radiatively stable. In~fact, we will look for scenarios for which the potential is SI along $\phi_1$, with $\phi_2 =0$, or {\it vice versa}, where $\phi_2$ is arbitrary, but the field $\phi_1$ is fixed at the origin. 

For definiteness, let us consider the second option mentioned above and write down the rather simple potential
\begin{equation}
    \label{eq:VU1sigma}
V_\phi (\phi_1\,,\phi_2)\ =\   m^2_{11}\, |\phi_1|^2\: +\: \big( m^2_{12}\, \phi^*_1 \phi_2 + {\rm c.c.}\big)\: +\: 
\lambda_1 |\phi_1|^4\: +\: 
\lambda_2 |\phi_2|^4\: ,
\end{equation}
which is SI in the $\phi_2$ direction when $\phi_1=0$. Notice that the interaction between $\phi_1$- and $\phi_2$-sectors enters only through dimension-2 terms proportional to $m^2_{12}$. In the limit of $m^2_{12}\to 0$, the model has a higher global symmetry described by the product group $\text{U}(1)_{\phi_1}\times \text{U}(1)_{\phi_2}$. Thus, a non-zero $m^2_{12}$ gives rise to the breaking pattern
\begin{equation}
    \label{eq:U1xU1}
\text{U}(1)_{\phi_1}\times \text{U}(1)_{\phi_2}\ \overset{m^2_{12}\ne 0}{\longrightarrow}\ \text{U}(1)_{\phi_1+\phi_2} \equiv\, \text{U}(1)\, . 
\end{equation}
Note that $\text{U}(1)_{\phi_1}\times \text{U}(1)_{\phi_2}$ does not forbid the portal term $\lambda_3\,|\phi_1|^2|\phi_2|^2$ in the potential~\eqref{eq:VU1sigma}. 

The only way to prohibit a portal term $\propto \lambda_3$ will be to require that the $\phi_2$-sector is SI, independently of the $\phi_1$-sector which can be scale-violating~\cite{Foot:2013hna}. Specifically, if $m^2_{11}$ is set to zero in addition to $m^2_{12}$, the global U(1) model will exhibit a larger group of scaling or dilatonic symmetries given by the product: 
$\text{D}_{\phi_1}\times \text{D}_{\phi_2}$, where $\text{D}_{\phi_1}$ ($\text{D}_{\phi_2}$) is the dilatonic symmetry in the $\phi_1$- ($\phi_2$)-sector. Then, in the presence of $\lambda_3\,|\phi_1|^2|\phi_2|^2$, the following breaking pattern arises:  
\begin{equation}
    \label{eq:D1xD1}
\text{D}_{\phi_1}\times \text{D}_{\phi_2}\ \overset{\lambda_3\ne 0}{\longrightarrow}\ \text{D}_{\phi_1+\phi_2} \equiv\, \text{D}_\phi\, . 
\end{equation}
If $m^2_{11}$ is non-zero, the scaling symmetry $\text{D}_{\phi_1}$
is explicitly broken by dimension-2 operators, but $\text{D}_{\phi_2}$ remains intact. Furthermore, if $m^2_{12}$ is switched on, this will generally break $\text{D}_{\phi_2}$ `softly', which can only be preserved in the $\phi_2$-field direction when $\phi_1 =0$. As a consequence of the above considerations, we have the following RGIs resulting from the potential~\eqref{eq:VU1sigma}:
\begin{equation}
    \label{eq:U1sigmaRGIs}
\beta_{m^2_{22}}\: =\: 0\,,\qquad \beta_{\lambda_3}\: =\: 0\,.     
\end{equation}
The vanishing of $\beta_{m^2_{22}}$ and $\beta_{\lambda_3}$ will still hold if Yukawa interactions are added to the theory, provided that neither $\phi_{1,2}$ couples to the same set of fermions. For the same reason, the U(1) scenario considered in this subsection cannot be promoted to a local model, because gauge interactions alone can generate the operator $|\phi_1|^2|\phi_2|^2$ at the one-loop level. On the~other~hand, the simple U(1) scenario we have been discussing here exemplifies the important role that scaling (dilatonic) symmetries can play in the identification of RGI quantities to all orders. 

\section{RG Invariants in Two-Higgs Doublet Models}\label{sec:2HDM}

We now turn our attention to a realistic extension of the SM, such as the 2HDM, with the aim of finding RGIs among the parameters in its scalar potential. As mentioned in the introduction, we should keep in mind that the 2HDM potential can generally possess 13 SU(2)$_L$-invariant global symmetries, of which 6 symmetries are also invariant under the gauge group U(1)$_Y$. In addition, if all bilinear mass terms are absent, the 2HDM will be SI in the Born approximation. 

Before discussing the crucial interplay between scaling and non-overlapping symmetries of the 2HDM, let us briefly review the tree-level form of its Lagrangian,
\begin{equation}
   \label{eq:L2HDM}
  {\cal L}\;  =\; (D_\mu\phi_1)^\dagger (D^\mu \phi_1)\, +\,
  (D_\mu\phi_2)^\dagger (D^\mu \phi_2)\, -\, V\, +\, {\cal L}_{\rm Y}\; . 
\end{equation}
In the above, $D_\mu= {\bf 1}_2\partial_\alpha + \frac{i}{2}g \sigma^i W^i_\alpha
+\frac{i}{2} g' {\bf 1}_2B_\alpha$ is the covariant spacetime
derivative with respect to the SM gauge group that acts on the two Higgs doublets~$\phi_1$ and~$\phi_2$ (with $\sigma^{i\,=\,1,2,3}$ denoting
the three Pauli matrices). Moreover, the scalar potential~$V$ of the theory is as follows:
\begin{eqnarray}
  \label{eq:V2HDM}
V \!& = &\! m_{11}^2 (\phi_1^{\dagger} \phi_1) + m_{22}^2
(\phi_2^{\dagger} \phi_2) + \Big[ m_{12}^2 (\phi_1^{\dagger} \phi_2) + {\rm H.c.}\Big] + \lambda_1 (\phi_1^{\dagger} \phi_1)^2 + \lambda_2 (\phi_2^{\dagger} \phi_2)^2 + \lambda_3 (\phi_1^{\dagger}
\phi_1)(\phi_2^{\dagger} \phi_2)\nonumber\\
\!&&\!  +\: \lambda_4 (\phi_1^{\dagger}
\phi_2)(\phi_2^{\dagger} \phi_1)  + \Big[\, \frac{\lambda_5}{2} (\phi_1^{\dagger} \phi_2)^2 + +\: \lambda_6 (\phi_1^{\dagger} \phi_1) (\phi_1^{\dagger} \phi_2) + 
\lambda_7 (\phi_2^{\dagger} \phi_2) (\phi_1^{\dagger} \phi_2) +
{\rm H.c.}\Big]\;.
\end{eqnarray}
Note that the general CP-violating 2HDM potential $V$~\cite{Pilaftsis:1999qt} contains 4 real mass parameters,
$m_{11}^2$, $m_{22}^2$, ${\rm Re}\,m_{12}^2$ and ${\rm Im}\,m^2_{12}$,
and 10 real quartic couplings, $\lambda_{1-4}$,
${\rm Re}\,\lambda_{5-7}$ and ${\rm Im}\,\lambda_{5-7}$.
Using\- the freedom of an SU(2) reparameterisation between the two Higgs doublets, one may eliminate three parameters at the classical level. Nevertheless, beyond the Born approximation
and depending on the adopted gauge-fixing scheme,
all  14~parameters may be necessary for the renormalisability of the theory in the {\em off-shell} kinematic region, such as
the effective potential${}$~\cite{Binosi:2005yk,Alexander:2008hd}. However, the $\beta$-functions of gauge-invariant parameters or operators do not depend on the gauge-fixing scheme.  

The last term of the Lagrangian ${\cal L}$~\eqref{eq:L2HDM}, ${\cal L_{\rm Y}}$, describes the Yukawa interactions of the Higgs doublets to quarks and leptons. If we consider only quark states for simplicity, the Yukawa Lagrangian ${\cal L_{\rm Y}}$ may be expressed as~\cite{BhupalDev:2014bir}
\begin{equation}
  \label{eq:LYuk}
- {\cal L}_{\rm Y}\ =\ \overline{Q}_L\, {\cal Y}_Q(\phi_i)\: Q_R\ +\ {\rm H.c.}\,, 
\end{equation}
where $Q_{L\,(R)} = ( u_{L\,(R)}\,, d_{L\,(R)})^{\sf T}$ and
\begin{equation}
   \label{eq:calMQ}
{\cal Y}_Q(\phi_i)\ =\ \left(\! \begin{array}{cc}
 h^u_i i\sigma^2 \phi^*_i    &  h^d_i \phi_i
\end{array}
\!\right)  
\end{equation}
is an SU(2)$_L$ gauge-covariant $\phi_i$-dependent matrix. In~\eqref{eq:calMQ}, the summation of the repeated index $i=1,2$ is implied. Moreover, in~writing ${\cal L_{\rm Y}}$ in~\eqref{eq:LYuk}, we suppressed the inter-family indices of the down- and up-quark Yukawa-coupling matrices,~$h^d_{1,2}$ and $h^u_{1,2}$, associated with the Higgs doublets $\phi_{1,2}$ and their hypercharge-conjugate doublets~$i\sigma^2 \phi^*_{1,2}$.

\begin{table}[t!]
    \centering
    \begin{tabular}{|l||c c c c c c c c c c|}
        \hline
Symmetry & ~$m^2_{11}$~ &  ~$m^2_{22}$~ & ~$m^2_{12}$~ & ~~$\lambda_1$~~ & ~~$\lambda_2$~~ & ~~$\lambda_3$~~ & ~~$\lambda_4$~~ & ~$\lambda_5$~ & ~~$\lambda_6$~~ & ~~$\lambda_7$~~
\\[0.1cm]
        \hline\hline
        U(1)$_{\rm PQ}$ & $-$ & $-$ & 0 & $-$ & $-$ & $-$ & $-$ & 0 & 0 & 0 \\
        \hline
        CP2 & $-$ & $m^2_{11}$ & 0 & $-$ & $\lambda_1$ & $-$ & $-$ & $-$  & $-$ & $-\lambda_6$\\
        \hline
        CP3 & $-$ & $m^2_{11}$ & 0 & $-$ & $\lambda_1$ & $-$ & $-$ & $2\lambda_1\! -\!\lambda_{34}$  & $0$ & $0$\\
        \hline\hline  
 $\text{U}(1)_{\rm PQ} \times \text{CP2}$ & $-$ & $m^2_{11}$ & 0 & $-$ & $\lambda_1$ & $-$ & $-$ & 0 & 0 & 0\\
 \hline
    \end{tabular}
\caption{\em U(1)$_{\rm PQ}$, CP2 and CP3 symmetry relations of kinematic parameters in the 2HDM. with $\lambda_{34}\equiv \lambda_3 + \lambda_4$. Note that $\text{CP3} \simeq  \text{U}(1)_{\rm PQ} \times \text{CP2}$ [cf.~\eqref{eq:CP3iso}].}
    \label{tab:2HDMParam}   
\end{table}

\subsection{CP2 Symmetric 2HDM}\label{subsec:2HDMCP2}

To find candidate RGIs, we follow the strategy developed in the previous section and look for scenarios that feature an SI field direction which does not get spoiled by UV divergences at the quantum loop level. To achieve this, we impose the CP2 symmetry, as stated in~\eqref{eq:CP2U1}, on the scalar potential~\eqref{eq:V2HDM}. This leads to the symmetry relations given in Table~\ref{tab:2HDMParam}. In close analogy with the observation made in the global U(1) model in~\eqref{eq:sigmadir} and~\eqref{eq:VU1m2},  it is not difficult to verify that along the SI field direction
\begin{equation}
   \label{eq:fielddir} 
\phi_{2\pm}\: =\: \pm\, i\, e^{-i\theta_{12}}\,\phi_{1\pm}\,,
\end{equation}
with $\theta_{12} = {\rm arg}(m^2_{12})$, the bilinear part of the 2HDM potential~\eqref{eq:V2HDM},
\begin{equation}
    \label{eq:V2HDMm2}
V_{m^2}(\phi_1\,,\phi_2)\ =\ m^2_{11}\, \big(\phi^\dagger_1\phi_1\, -\, \phi^\dagger_2\phi_2\big)\: +\: |m^2_{12}|\,
\big( e^{i\theta_{12}} \phi^\dagger_1 \phi_2 + e^{-i\theta_{12}} \phi^\dagger_2\phi_1\big)\, ,   
\end{equation}
vanishes identically. This consideration forces us to impose the following restriction between the bilinear mass parameters: 
\begin{equation}
   \label{eq:2HDMconstr}
m^2_{22}\: =\: -m^2_{11}\,,
\end{equation}
while $m^2_{12}\ne 0$ can be arbitrary and complex. As first noted in~\cite{Ferreira:2023dke}, the restriction~\eqref{eq:2HDMconstr} violates the CP2 symmetry, but only softly, that is, by dimension-2 operators.  

\begin{table}[t!]
    \centering
    \begin{tabular}{|c||c c  |}
        \hline
Fields & PQ charge & CP2 parity \\[1mm]
        \hline\hline
$\phi_1\: (\phi_2)$ & $+1\: (-1)$ &   \\
        \hline
$u_{L,R}\,, d_{L,R}$ & $0$ &   \\
\hline
$\phi^\dagger_1\phi_1 \pm \phi^\dagger_2\phi_2$ & $0$ & $\pm 1$   \\
\hline
$\phi^*_1\phi_2\: (\phi^*_2\phi_1)$ & $-2\: (+2)$ & $-1$\\
        \hline\hline
Spurions & PQ charge & CP2 parity \\[1mm]
\hline\hline
$g^2_s\,,\, g^2\,,\, g'^2$ & $0$ & $+1$   \\
\hline
$m^2_{11} \pm m^2_{22}$ & $0$ & $\pm 1$   \\
\hline
$m^2_{12}\, (m^{2\,*}_{12})$ & $+2\: (-2)$ & $-1$   \\
\hline
$\lambda_1\pm \lambda_2$ & $0$ & $\pm 1$   \\
\hline
$\lambda_3\,,\lambda_4$ & $0$ & $+1$   \\
\hline
$\lambda_5\, (\lambda^*_5 )$ & $+4\: (-4)$ & $+1$   \\
\hline
$\lambda_6\pm \lambda_7$ & $+2$ & $\mp 1$   \\
\hline
$\lambda^*_6\pm \lambda^*_7$ & $-2$ & $\mp 1$   \\
\hline
$\text{Tr}\, [(h^{u}_{1(2)} h^{u\dagger}_{2(1)})\, \cdots ]$ & $+2\: (-2)$ & $-1$   \\
\hline
$\text{Tr}\, [ (h^{u}_1 h^{u\dagger}_1\pm h^{u}_2 h^{u\dagger}_2)\, \cdots ]$ & $0$ & $\pm 1$   \\
\hline
$\text{Tr}\, [(h^{d}_{1(2)} h^{d\dagger}_{2(1)})\, \cdots ]$ & $-2\: (+2)$ & $-1$   \\
\hline
$\text{Tr}\, [ (h^{d}_1 h^{d\dagger}_1\pm h^{d}_2 h^{d\dagger}_2)\, \cdots ]$ & $0$ & $\pm 1$   \\
\hline
    \end{tabular}
 \caption{\em PQ and CP2 charges of fields or field bilinears, and the associated spurion charges of parameter expressions in the 2HDM. The dots within the traces indicate Yukawa-coupling expressions with zero PQ charge and CP2 parity $+1$. For details, see discussion in the text.}
    \label{tab:2HDMspurion}
\end{table}

Our next task is to show that $\beta_{m^2_{11} + m^2_{22}} = 0$ in the 
softly broken CP2 symmetric 2HDM, under the restriction~\eqref{eq:2HDMconstr}. To this end, we employ spurion techniques that were developed in Section~\ref{sec:Sym}~[cf.~\eqref{eq:SGspurions}]. Table~\eqref{tab:2HDMspurion} gives the PQ charges and CP2 parities for all relevant fields and spurion expressions. As can be seen in this table, the gauge interactions of the strong and electroweak forces, mediated with gauge couplings $g_s$, $g$ and $g'$, do not carry spurion PQ charges and have even CP2 parity, and as such they cannot modify the RG structure of $\beta_{m^2_{11} + m^2_{22}}$. Likewise, the quartic interactions of the scalar potential~\ref{eq:V2HDM} do not alter the RG property: $\beta_{m^2_{11} + m^2_{22}} = 0$ either, for the same reasons as stated in the discussion of~\eqref{eq:betaU1}. The only addition is the coupling $\lambda_4$, but this has a zero spurion PQ charge and even CP2 parity, and, like $\lambda_3$ in the global U(1) model,  the vanishing of $\beta_{m^2_{11} + m^2_{22}}$ will still persist.

From the above discussion, it becomes clear that the only non-trivial effects could come from the Yukawa sector~\eqref{eq:LYuk}. As shown in~\cite{Maniatis:2007vn,Maniatis:2007de}, at least two generations of quarks and leptons are needed to get non-zero Yukawa couplings in a CP2-symmetric 2HDM. For the sake of illustration, let us therefore consider a two-generation model of quarks whose CP2-symmetric Yukawa matrices are 
\begin{equation}
    \label{eq:YukCP2}
h^q_1\: =\: 
\left(\! \begin{array}{cc}
  a_q   &  b_q \\
  b_q   &  -a_q
\end{array}\!\right) =\: a_q\sigma^3 + b_q\sigma^1 \,,\quad
h^q_2\: =\: 
\left(\!\! \begin{array}{cc}
  -\,b^*_q   &  a^*_q \\
  a^*_q   &  b^*_q
\end{array}\!\right) =\: -b^*_q\sigma^3 + a^*_q\sigma^1\,,
\end{equation}
with $q=u,d$, and $a_q,b_q\in \mathbb{C}$. Since the right-handed quarks $u_R$ and $d_R$ will appear as propa\-gators within loops under the action of a trace, the relevant spurion expressions are easily found to be
\begin{eqnarray}
    \label{eq:hq1221}
h^q_1 h^{q\dagger}_2 \!&=&\! i \big(a^2_q + b^2_q\big)\sigma^2\,,\qquad  h^q_2 h^{q\dagger}_1\: =\: -i \big(a^{*2}_q + b^{*2}_q\big)\sigma^2\,,\\
    \label{eq:hq1122}
h^q_1 h^{q\dagger}_1 +(-)\: h^q_2 h^{q\dagger}_2\!&=&\! 2\big(|a_q|^2 + |b_q|^2\big){\bf 1}_2\quad \Big(2i\big(a_q b^*_q - b_q a^*_q\big)\sigma^2 \Big)\,,
\end{eqnarray}
where $\sigma^2$ is the second Pauli matrix and ${\bf 1}_2$ is the 2D unit matrix. With the exception of the CP2-even combination $h^q_1 h^{q\dagger}_1 +\, h^q_2 h^{q\dagger}_2 \propto {\bf 1}_2$ in~\eqref{eq:hq1122}, all other CP2-odd spurion expressions, as well as odd powers of these, vanish when the trace is taken, such as $\text{Tr}\,[(h^q_1 h^{q\dagger}_2)^{2k+1}] = 0$, with $k=0,1,2,\dots$ However, even powers of CP2-odd spurion expressions are proportional to~${\bf 1}_2$ and therefore they are not traceless, e.g.~$\text{Tr}\,[(h^q_1 h^{q\dagger}_2)^{2k}] = 2\,(-1)^k (a^2_q + b^2_q)^{2k}$, etc.  
Finally, we should observe that all Yukawa-matrix spurion expressions in~\eqref{eq:hq1221} and~\eqref{eq:hq1122} commute with each other. 

With the aid of Table~\ref{tab:2HDMspurion}, we may now examine the structure of $\beta_{m^2_{11} + m^2_{22}}$ to all orders. Like in~\eqref{eq:betaU1}, we may write its general form as
\begin{equation}
    \label{eq:beta2HDM}
\beta_{m^2_{11}+m^2_{22}}\: =\: A\, (m^2_{11} + m^2_{22})\, +\, B\, (m^2_{11} - m^2_{22})\, +\, \big( C\, m^2_{12}\: +\: \text{c.c.}\big)\; .
\end{equation}
If we look for Yukawa-coupling contributions only to $B$ and $C$, these can only come from 
\begin{equation}
   \label{eq:betaBC}
    B\: \propto\: \text{Tr}\, [(h^{q}_1 h^{q\dagger}_1 - h^{q}_2 h^{q\dagger}_2)\dots]\,,\quad
    C\:\propto\:  \text{Tr}\, [(h^{u}_{2} h^{u\dagger}_{1})\, \cdots ]\,,\
    \text{Tr}\, [(h^{d}_{1} h^{d\dagger}_{2})\, \cdots ]\,,
\end{equation}
where the ellipses denote Yukawa-matrix spurion expressions with zero PQ charge and CP2 parity~$+1$. However, as discussed above, all the traces in~\eqref{eq:betaBC} evaluate to zero. The inclusion of quartic and gauge couplings, $\lambda_{1-7}$ and $g_s$, $g$ and $g'$, cannot change this result, yielding $B = C= 0$. Therefore, we may safely conclude that it is $\beta_{m^2_{11} + m^2_{22}} = 0$ to all orders in perturbation theory, within a softly broken CP2 symmetric 2HDM constrained by~\eqref{eq:2HDMconstr}.

It is instructive to examine whether the constraint $\lambda_1 - \lambda_2 =0$
is preserved beyond the tree level, i.e. whether $\beta_{\lambda_1 - \lambda_2} = 0$
to all orders. Since spurion $\lambda_1 - \lambda_2$ has an odd CP2 parity and zero PQ charge~[cf.~Table~\ref{tab:2HDMspurion}], its $\beta$-function can be decomposed as follows:
\begin{equation}
    \label{eq:betalambda12}
\beta_{\lambda_1 - \lambda_2}\: =\: a\, (\lambda_1 - \lambda_2)\: +\: b\, \text{Tr}\, [(h^{q}_1 h^{q\dagger}_1 - h^{q}_2 h^{q\dagger}_2)\dots]\,, 
\end{equation}
where $a$ and $b$ are dimensionless polynomials of quartic, Yukawa and gauge couplings that have vanishing spurion PQ charge and a spurion CP2 parity $+1$. Given the tree-level symmetry relations in~Table~\ref{tab:2HDMspurion} and in~\eqref{eq:hq1122}, we find $\beta_{\lambda_1 - \lambda_2}= 0$ to all orders, as should be in the CP2-symmetric 2HDM. By analogy, following a similar line of arguments, it is not difficult to show with the help of Table~\ref{tab:2HDMspurion} that $\beta_{\lambda_6 +\lambda_7} = 0$ to all orders.    

\subsection{CP3 Symmetric 2HDM}\label{subsec:2HDMCP3}

Another interesting 2HDM realisation with a global symmetry larger than CP2 is the CP3-symmetric 2HDM. According to~\cite{Ivanov:2006yq,Maniatis:2007de}, this scenario can be obtained by requiring
invariance of the classical action under the following field transformations:
\begin{equation}
    \label{eq:CP3}
\text{CP3:}\quad \left(\! \begin{array}{c}
    \phi_1 \\
    \phi_2  
\end{array}\!\right)\quad \to\quad 
\left(\! \begin{array}{c}
    \phi'_1 \\
    \phi'_2  
\end{array}\!\right)\: \equiv\: R(\theta ) \left(\! \begin{array}{c}
    \phi^*_1 \\
    \phi^*_2  
\end{array}\!\right)\: =\:  
\left(\!\begin{array}{cc}
\cos\theta & \sin\theta \\
-\sin\theta & \cos\theta    
\end{array}\!\right)
\left(\! \begin{array}{c}
    \phi^*_1 \\
    \phi^*_2  
\end{array}\!\right)\, ,
\end{equation}
with $R(\theta )\in \text{O}(2)$.
Notice that for $\theta = \pi/2$, $\text{CP3} \equiv \text{O}(2)\times \text{CP1}$ explicitly breaks to CP2, where $\text{CP1}$ is the standard CP transformation of fiels: $\phi_{1(2)} \to \phi'_{1(2)} = \phi^*_{1(2)}$. Thus, CP2 is contained within~$\text{CP3}$, i.e.~$\text{CP2}\subset   \text{CP3}$.  The parameter relations governing a scalar potential invariant under CP3 are given in Table~\ref{tab:2HDMParam}. We should also remark that a non-vanishing CP3-invariant Yukawa sector
cannot be realised, unless the custodial symmetric limit $g'\to 0$ is considered, according to our findings below.   

To be able to employ the spurion formalism for the CP3 scenario, we go to a field basis, where the group isomorphism 
\begin{equation}
  \label{eq:CP3iso}
 \text{CP3} \equiv \text{O}(2) \times \text{CP1}\: \simeq\: \text{U}(1)_{\rm PQ} \times \text{CP2}   
\end{equation}
is exploited. In other words, the CP3 symmetry is constrained by an extra $\text{U}(1)_{\rm PQ}$ in addition to CP2. To make this explicit, we use the harmonic coordinate field basis
\begin{equation}
   \label{eq:PQbasis}
 \left(\! \begin{array}{c}
    \phi_+ \\
    \phi_-  
\end{array}\!\right)\: \equiv\: \left(\! \begin{array}{c}
    \frac{1}{\sqrt{2}}\,(\phi_1 - i\phi_2 )\\
    \frac{1}{\sqrt{2}}\,(\phi_1 + i\phi_2 ) 
\end{array}\!\right)\: =\:  V\, \left(\! \begin{array}{c}
    \phi_1 \\
    \phi_2  
\end{array}\!\right)\: =\:  
\frac{1}{\sqrt{2}}\left(\!\begin{array}{cc}
1 & -i \\
1 & i    
\end{array}\!\right)
\left(\! \begin{array}{c}
    \phi_1 \\
    \phi_2  
\end{array}\!\right)\, .   
\end{equation}
Under $\text{U}(1)_{\rm PQ} \times \text{CP2}$, the newly introduced fields transform as
\begin{equation}
    \label{eq:PQCP2}
\left(\! \begin{array}{c}
    \phi'_+ \\
    \phi'_-  
\end{array}\!\right)\: =\: D(\theta )\, i\sigma^2\, \left(\! \begin{array}{c}
    \phi^*_+ \\
    \phi^*_-  
\end{array}\!\right)\: =\: \left(\!\begin{array}{cc}
e^{i\theta} & 0 \\
0 & -e^{-i\theta}    
\end{array}\!\right)
\left(\!\begin{array}{cc}
0 & 1 \\
-1 & 0    
\end{array}\!\right)
\left(\! \begin{array}{c}
    \phi^*_+ \\
    \phi^*_-  
\end{array}\!\right)\, .
\end{equation}
As a consequence of $\text{U}(1)_{\rm PQ} \times \text{CP2}$, the  term $(\phi^\dagger_+ \phi_-)^2$ disappears from the scalar potential, since in this field basis the scalar doublets $\phi_+$ and $\phi_-$ have the well defined PQ charges $+1$ and~$-1$, respectively. However, as can be seen in Table~\ref{tab:2HDMParam}, the number of independent parameters remains the same, as should
be. 

Given the definite PQ charges of $\phi_\pm$, only a Type-II 2HDM realisation for the Yukawa sector can be made compatible with CP2. In detail, the Lagrangian describing the quark Yukawa interactions is given by
\begin{equation}
    \label{eq:CP3Yukawa}
-\,{\cal L}_{\text{Y}}\: =\: \overline{Q}_L h^u \widetilde{\phi}_+ u_R\: +\:    \overline{Q}_L h^d \phi_- d_R\  +\ {\rm H.c.}\,, 
\end{equation}
with $\widetilde{\phi}_\pm = i\sigma_2\phi^*_\pm$. The imposition of CP2 on the scalar doublets in~\eqref{eq:PQCP2} and the quark fields, 
\begin{eqnarray}
    \label{eq:CP2quarks}
 u'_R\: =\: d^C_R\,,\qquad d'_R\: =\: -u^C_R\, ,   
\end{eqnarray}
implies the quark Yukawa matrix equality
\begin{equation}
    \label{eq:huhd}
h^u\: =\: h^{d*}\, .     
\end{equation}
Observe that this equality is only possible in the custodially symmetric limit of the theory, where U(1)$_Y$ gauge interactions are switched off by setting $g' =0$. The relevant spurion expressions for the Yukawa matrices are
\begin{equation}
    \label{eq:CP3spurion}
    {\rm Tr}\,\big[\big(h^u h^{u\dagger} + h^d h^{d\dagger}\big)\dots ]\,,\qquad {\rm Tr}\,\big[\big(h^u h^{u\dagger} - h^d h^{d\dagger}\big)\dots ]\,,
\end{equation}
which have zero spurion PQ charges and CP2 partities $+1$ and $-1$, respectively. 
As before, we find $\beta_{m^2_{11}+m^2_{22}} = 0$ (with the constraint~\eqref{eq:2HDMconstr}) and $\beta_{\lambda_1 - \lambda_2} = 0$ to all orders, in the chiral anomaly free limit $g'\to 0$. If $g'\ne 0$, the equality $\lambda_1 = \lambda_2$ is violated
by two-loop diagrams that mediate U(1)$_Y$ gauge bosons $B_\mu$, e.g.~between two $u_R$ quark lines or between two $d_R$ quark lines. In this case, one finds that the RG $\beta$-function
\begin{equation}
   \label{eq:CP3lambda12}
 \beta_{\lambda_1 - \lambda_2}\: \simeq\: \frac{g'^2}{(16\pi^2)^2} \Big[\,   y^2_d\, {\rm Tr}\, (h^d h^{d\dagger})^2\: -\: y^2_u\,  {\rm Tr}\, (h^u h^{u\dagger})^2\,\Big]\: +\: \dots
\end{equation}
is non-zero under the constraint~\eqref{eq:huhd}, because
$u_R$ and $d_R$ have different weak hypercharges~$y_{u,d}$, i.e.~$y_u = 4/3$ and $y_d = -2/3$. The ellipses in~\ref{eq:CP3lambda12} denote other possible contributions at two loops, e.g.~from an exchange of a $B_\mu$ gauge boson between a doublet $Q_L$ line and a $u_R$ or $d_R$ quark in the loop. Consequently, {\em exact} CP3-symmetric 2HDM scenarios cannot be realised when $g' \ne 0$ {\em and} $h^{u,d} \ne 0$.

\subsection{The Gauge Hierarchy Problem beyond Supersymmetry}\label{subsec:SUSY}

An important theoretical motivation to look for supersymmetric theories originates from their technical ability to solve the infamous gauge hierarchy problem~\cite{Gildener:1976ai}. To put it simply, the electroweak scale as determined by the SM usually gets destabilised by the potential presence of high mass scales in the theory of order GUT scale through quantum loop effects. It can usually be driven to higher values close to the high scale, unless a very fine-tuned cancellation between the electroweak scale and high mass scale contributions is considered. Super\-symmetry solves this problem technically thanks to non-renormalisation theorems that govern the superpotential of supersymmetric theories~\cite{Martin:1997ns}, including the so-called Minimal Supersymmetric Standard Model~\cite{Djouadi:2005gj}.
If a dimensionful parameter is absent from the superpotential, this parameter cannot reappear radiatively by perturbative quantum effects to all orders. 

In nature, supersymmetry needs to be broken, at least softly, for theoretical and pheno\-menological reasons. If this soft breaking is at the TeV scale, the aforementioned fine-tuning among different quantum contributions will not be excessive. On the other hand, the non-observation of supersymmetry in the current LHC data puts a lower limit on the soft-super\-symmetry breaking scale of order 10~TeV, thereby inducing a tension in cancelling contributions from SM particles and their supersymmetric counterparts to the Higgs-boson mass. This tension is often called the little hierarchy problem~\cite{Barbieri:2000gf}. 

As was already mentioned in Section~\ref{sec:RGIs}, classical scaling symmetries may be regarded as another valid possibility to solve the gauge hierarchy problem. However, explicit breaking of these symmetries through the presence of any high-scale masses results in reintroducing the gauge hierarchy problem, unless there is a miraculous no mixing mechanism between scales at quantum level in the so-called {\em Higgs basis}~\cite{GEORGI197995,Lavoura:1994fv,Davidson:2005cw}, in which $\langle \phi_2\rangle = 0$ but $\langle \phi_1\rangle \neq 0$. Here, we would like to demonstrate that a softly broken CP2-symmetric 2HDM can achieve such a profound separation of mass scales due to the synergy of scaling and non-overlapping symmetries, such as U(1)$_{\rm PQ}$ and CP2. However, in this section, we will argue that this separation is not sufficient to address the gauge-hierarchy problem within perturbation theory.

Using the spurion charge assignments from Table~\ref{tab:2HDMspurion}, it is straightforward to show in the field basis $\lambda_6 = \lambda_7 \ne 0$ that
\begin{eqnarray}
    \label{eq:RGscales}
    \beta_{m^2_{11} + m^2_{22}} \!&=&\! A_+\,(m^2_{11} +m^2_{22})\: +\: B_+\,(\lambda_1 -\lambda_2)\,(m^2_{11} -m^2_{22})\nonumber\\
    \!&&\!\: +\: \Big\{\Big[ C^{(1)}_+\,(\lambda^*_6 +\lambda^*_7) +   C^{(2)}_+\, \lambda^*_5(\lambda_6 + \lambda_7)\Big]\, m^2_{12}\ +\ {\rm c.c.}\Big\}\,,\nonumber\\[2mm]
    \beta_{m^2_{11} - m^2_{22}} \!&=&\! A_-\,(m^2_{11} -m^2_{22})\: +\: 
    B_-\,(\lambda_1 -\lambda_2)\,(m^2_{11} +m^2_{22})\nonumber\\
     \!&&\!+\: \Big[  C_-\, (\lambda_1 -\lambda_2) (\lambda^*_6 + \lambda^*_7)\, m^2_{12}\ +\ {\rm c.c.}\Big]\,,\\[2mm]
    \beta_{m^2_{12}} \!&=&\! A_{12}\,m^2_{12}\: +\: \big[ B^{(1)}_{12}\,\lambda_5 + B^{(2)}_{12} (\lambda_6 +\lambda_7)^2 \big]\, m^{2\,*}_{12}\nonumber\\ 
    \!&&\!
    \: +\:\Big[ C_{12}\, (\lambda_1 -\lambda_2) (\lambda_6 + \lambda_7)\,(m^2_{11} - m^2_{22})\ +\ {\rm c.c.}\Big]\,,\nonumber
\end{eqnarray}
where the prefactors $A_\pm$, $A_{12}$, $B_+$ {\it etc} are functions of gauge, quartic and Yukawa couplings with no spurion charge. In deriving~\eqref{eq:RGscales}, we did not resort to the parameter relations in Table~\ref{tab:2HDMParam} and the constraint  in~\eqref{eq:2HDMconstr}, but only assumed, for simplicity, the CP2-symmetric Yukawa matrices in~\eqref{eq:YukCP2}, in the field basis in which $\lambda_6 =\lambda_7 \ne 0$~\footnote{We note that the consideration of an arbitrary field basis where $\lambda_6 \ne \lambda_7 \ne 0$ only proliferates the number of independent spurion polynomials of quartic couplings without invalidating the proof presented here.}. As a consequence, {\em non}-zero trace expressions of quark Yukawa matrices can only have zero spurion charge and as such, they can only appear in the prefactors $A_\pm$, $A_{12}$, $B_+$ {\it etc}. Interestingly enough, the three mass scales of the theory, $m^2_{11} \pm m^2_{22}$ and~$m^2_{12}$, run independently under RG equations, provided that we set $\lambda_1 = \lambda_2$ and $\lambda_6 = -\lambda_7$ in~\eqref{eq:RGscales}. 

Based on our spurion formalism and applying it in the field basis $\lambda_6 =\lambda_7\ne 0$, we may also derive the structure of RG $\beta$-functions to all orders:
\begin{eqnarray}
    \label{eq:RGlambdas}
\beta_{\lambda_1 - \lambda_2} \!&=&\! a_-\, (\lambda_1 - \lambda_2)\,,\nonumber\\
\beta_{\lambda_5} \!&=&\! a_5\, \lambda_5\: +\:  b_5\, (\lambda_6 +\lambda_7)^2\,,\nonumber\\
\beta_{\lambda_6 + \lambda_7} \!&=&\! a^+_{67}\, (\lambda_6 + \lambda_7)\: +\: b^+_{67}\, \lambda_5(\lambda^*_6 + \lambda^*_7)\,,\\
\beta_{\lambda_6 - \lambda_7} \!&=&\!   b^-_{67}\, (\lambda_1 -\lambda_2) (\lambda_6 + \lambda_7)\,.\nonumber
\end{eqnarray}
Like in~\eqref{eq:RGscales}, the coefficients $a_-$, $a_5$, $b_5$ {\it etc} are  functions that depend on the dimensionless couplings of the theory and have no spurion charge, i.e.~their PQ charge and CP2-parity is 0 and $+1$, respectively. From~\eqref{eq:RGlambdas}, we readily see that the relations $\lambda_1 = \lambda_2$ and $\lambda_6 \pm \lambda_7 = 0$ are RGIs to all orders in perturbation theory, provided that the Yukawa sector is CP2 symmetric.

It is beyond the scope of the present work to address the phenomenological viability of the softly broken CP2-symmetric 2HDM under consideration. Strictly speaking, three-generation quark Yukawa matrices invariant under CP2 are not admissible in the theory~\cite{Maniatis:2007de}. Although an exact CP2 symmetric Yukawa sector may not be allowed when three generations are considered, it will still be worth investigating whether small departures from the CP2 symmetry lead to interesting phenomenology. In fact, by appropriately selecting the small deviations of the $3\times 3$ quark Yukawa matrices, the breaking of the CP2 symmetry can in principle be postponed to sufficiently higher loops.

To give a simple workable example, let us imagine that the two commuting 2D flavour matrices  $\sigma^2$ and ${\bf 1}_2$ in~\eqref{eq:hq1221} and~\eqref{eq:hq1122} could be extended and replaced with the 3D flavour matrices $A$ and~$S$, respectively, such that
\begin{equation}
    \label{eq:AqSq}
    A^{\sf T}\: =\: - A\,,\qquad  S^{\sf T}\: =\: S\, .
\end{equation}
We may also assume that the complex 3D matrices $S$ and $A$ do not commute with each other, i.e.~$[S\,, A] \ne 0$. Then, the first  Yukawa expression with non-vanishing trace and spurion charge is given by a structure of the form
\begin{equation}
    \label{eq:SA6loops}
    \text{Tr}\, \big(SAS^2A^2\big)\: =\: -\text{Tr}\, \big( SA^2S^2A\big)\,,
\end{equation}
since $SAS^2A^2 \ne SA^2S^2A$ in general. Such a structure could arise at 6 loop order, if one counts the necessary $u_R$ or $d_R$ quark propagators required in a higher order quantum loop. Now, counting the loop factors $(16\pi^2)^{-n} \simeq (1.6\times 10^2)^{-n}$, we get an enormous suppression factor of order~$10^{-13}$ for $n=6$, leading to unobservable deviations in the Higgs self-couplings from the exact CP2-symmetric 2HDM.

One may question whether the present spurion approach to finding RGIs that lead to separation of mass scales can provide an interesting alternative to technically solve the gauge hierarchy problem. To see that this is not generally the case, let us assume for simplicity that the dimension-4 part of the 2HDM potential is CP3 symmetric, with $\lambda_1 = \lambda_2 = (\lambda_3 + \lambda_4 +\lambda_5)/2$, while $\text{Im}(m^2_{12}) =0$. We note that this is a 2HDM scenario that features natural alignment~\cite{BhupalDev:2014bir,Darvishi:2023fjh}. As a consequence, we can diagonalise the bilinear part of the scalar potential by an O(2) rotation, without altering the quartic couplings and hence their relations. In this way, we can go to the scalar doublets in the Higgs basis,
\begin{equation}
    \label{eq:Hbasis}
\left(\! \begin{array}{c}
\widehat{\phi}_1\\
\widehat{\phi}_2
\end{array}\!\right)\: =\: R(\beta)\, \left(\! \begin{array}{c}
\phi_1\\
\phi_2
\end{array}\!\right) ,     
\end{equation}
where the two-by-two matrix $R(\beta)\in \text{O}(2)$ is determined through
\begin{equation}
   \label{eq:Masses_Hbasis}
R(\beta )\left(\! \begin{array}{cc}
m^2_{11} & \text{Re}\,m^2_{12}\\
\text{Re}\,m^2_{12} & m^2_{22}
\end{array}\!\right) R^{\sf T}\!(\beta) \: =\: \left(\! \begin{array}{cc}
\widehat{m}^2_{11} & 0\\
0 & \widehat{m}^2_{22}
\end{array}\!\right)\,,
\end{equation}
with $\tan2\beta = 2\,\text{Re}\,m^2_{12}/(m^2_{22} - m^2_{11})$ for $|\widehat{m}^2_{11}| \le |\widehat{m}^2_{22}|$.
To trigger the Higgs mechanism in the Higgs basis, $\widehat{m}^2_1$ must be negative, whereas $\widehat{m}^2_2$ should be positive, so that $\langle \widehat{\phi}_1 \rangle = (0\,,v_{\rm SM}/\sqrt{2})^{\sf T}$ and  $\langle \widehat{\phi}_2 \rangle = 0$, where $v_{\rm SM}\simeq 246$~GeV is the SM vacuum expectation value. 

In the Higgs basis as defined from~\eqref{eq:Hbasis} and~\eqref{eq:Masses_Hbasis}, it is not difficult to convince ourselves that a possible mass-scale hierarchy, $|\widehat{m}^2_1| \ll \widehat{m}^2_2$, is not protected, already at the one-loop quantum level. In the DR-$\overline{\rm MS}$ scheme (also known as $\overline{\rm DR}$ scheme) of renormalisation, the one-loop finite radiative correction to $\widehat{m}^2_{11}$ at the RG scale $Q^2 = v^2_{\rm SM}$ is given by 
\begin{equation}
    \label{eq:dhatm211}
\delta \widehat{m}^2_{11}\: \simeq\: -\, \frac{6\lambda_1}{16\pi^2}\, 
\widehat{m}^2_{11}\, \ln\bigg(\frac{|\widehat{m}^2_{11}|}{v^2_{\rm SM}}\bigg)\, -\:  \frac{2\lambda_3 + \lambda_4}{16\pi^2}\, 
\widehat{m}^2_{22}\, \ln\bigg(\frac{\widehat{m}^2_{22}}{v^2_{\rm SM}}\bigg)\,.
\end{equation}
Evidently, up to a loop factor of order $10^{-3}$, one-loop effects drive the value of  $\widehat{m}^2_{11}$ to the higher mass scale $\widehat{m}^2_{22}$. If $\widehat{m}_{22} \sim 10^{16}$~GeV is of the order of the scale of Grand Unification Theory (GUT), then the finite shift to $\delta \widehat{m}_{11}$ is of the order of $10^{14}$~GeV, which would destabilise the electroweak scale unless a UV-finite counter-term is introduced to finely cancel this enormous contribution. Hence, the softly-broken CP3-symmetric 2HDM suffers from the gauge-hierarchy problem. A similar conclusion can be reached for the softly-broken CP2-symmetric 2HDM, if~analysed in the Higgs basis. 

In light of the above discussion, the only alternative to the gauge-hierarchy problem would be to impose scale invariance on the 2HDM~\cite{Lee:2012jn,Eichten:2022vys,Slavich:2026zmi}, where 
$\beta_{m^2_{11} \pm m^2_{22}} = \beta_{m^2_{12}} = 0$ are all automatically fulfilled. However,
the presence of a Landau pole in the TeV region implies that the  SI-2HDM is on the verge of being excluded. To avoid this TeV-scale
Landau pole, one must consider further extensions or supersymmetric embeddings of the SI-2HDM. Such an analysis lies beyond the scope of this paper.

Finally, going beyond the 2HDMs to multi-HDMs~\cite{deMedeirosVarzielas:2011zw,Darvishi:2019dbh,Kuncinas:2025uty}, the situation becomes more laborious in terms of the multitude of SI field directions that may be present and for finding the synergetic properties of all non-overlapping symmetries. In spite of this technical complexity, the spurion formalism developed in this work provides a rigorous methodology to explore the separation of all different mass scales in models with an extended Higgs-scalar sector, thereby offering a tractable way to identify new RGIs and study their phenomenological implications.

\section{Conclusions}\label{sec:concl}

Renormalisation Group Invariants are essential features of QFTs with underlying global or local symmetries. Sometimes these symmetries may be hidden or partially broken, and can still be the origin of unexpected RGIs in new physics theories. In SUSY, non-renormalisation theorems, together with their soft breaking, give a rigorous theoretical framework to establish stable RGI fixed points.  

In this work, we go beyond SUSY and present a different approach to looking for RGIs. We~have shown how the synergy between scaling and non-overlapping symmetries can be responsible for RGIs among dimensionful parameters of extended scalar potentials beyond the~SM. To this end, we have employed the spurion-field formalism, which was developed further on the basis of~\eqref{eq:SGspurions}, to include models with extended Higgs-scalar potentials, such as the 2HDM. After illustrating how our spurion-field approach works in simple U(1) models in Section~\ref{sec:RGIs}, we have applied this approach to softly broken CP2-symmetric 2HDMs, where the charges of all relevant spurions are given in Table~\ref{tab:2HDMspurion}. In particular, we have found that the absence of the operator $r_0\equiv \phi^+_1\phi_1 + \phi^+_2\phi_2$ in the bilinear field part of the 2HDM potential may be reinforced by requiring a scaling symmetry along the SI field direction~\eqref{eq:fielddir}. In addition, the synergy of the non-overlapping CP2 and PQ symmetries ensures that this operator vanishes to all orders in perturbation theory, according to our derivation using the spurion-field formalism in Section~\ref{subsec:2HDMCP2}. In addition, 
using the spurion formalism resulting from the group isomorphism~\eqref{eq:CP3iso},  we have shown in Section~\ref{subsec:2HDMCP3} that an exact CP3-invariant 2HDM scenario cannot be realised if $g' \ne 0$ and $h^{u,d} \ne 0$. 

The spurion-field formalism presented in this paper enables us to
understand the separation of all mass scales in a theory under RG running. As we have shown in Section~\ref{subsec:SUSY}, such a separation can indeed take place in the softly broken CP2-symmetric 2HDM, in which the three mass scales $m^2_{11}\pm m^2_{22}$ and $m^2_{12}$ run independently under the RG equations~\eqref{eq:RGscales}. Our spurion-field approach has given new insight into the required structure for an approximate CP2-symmetric Yukawa sector that could be compatible with phenomenological constraints. However, we have shown that this is not sufficient to avoid destabilising the electroweak scale, and further extensions or embeddings of the model are necessary. It would be interesting to analyse these possibilities in future work. Another research direction would be to investigate whether a sufficient number of scaling and non-overlapping symmetries exists to lead to RGIs and allow for separation of all mass scales in theories of new physics beyond the 2HDM.

\subsection*{Acknowledgements} The author wishes to thank the reviewer of this article for enlightening comments regarding the gauge-hierarchy problem. This work is supported in part by
the STFC research grant: ST/X00077X/1.

\vfill\eject
\bibliography{RGIs}

\end{document}